# Development of a persuasive User Experience Research (UXR) Point of View for Explainable Artificial Intelligence (XAI)

Mohammad Naiseh
Computing and Informatics
Bournemouth University
Poole, UK
mnaiseh1@bournemouth.ac.uk

Huseyin Dogan
Computing and Informatics
Bournemouth University
Poole, UK
hdogan@bournemouth.ac.uk

Stephen Giff
Google
United States
sgiff@google.com

Nan Jiang
Computing and Informatics
Bournemouth University
Poole, UK
njiang@bournemouth.ac.uk

## ABSTRACT
Explainable Artificial Intelligence (XAI) plays a critical role in fostering user trust and understanding in AI-driven systems. However, the design of effective XAI interfaces presents significant challenges, particularly for UX professionals who may lack technical expertise in AI or machine learning. Existing explanation methods, such as SHAP, LIME, and counterfactual explanations, often rely on complex technical language and assumptions that are difficult for non-expert users to interpret. To address these gaps, we propose a UX Research (UXR) Playbook for XAI — a practical framework aimed at supporting UX professionals in designing accessible, transparent, and trustworthy AI experiences. Our playbook offers actionable guidance to help bridge the gap between technical explainability methods and user-centred design, empowering designers to create AI interactions that foster better understanding, trust, and responsible AI adoption.

## 1 Introduction

Explainable Artificial Intelligence (XAI) is a crucial component of modern AI systems, aiming to make model predictions understandable and interpretable to users [1]. However, designing effective and engaging XAI interfaces remains a significant challenge, particularly for UX researchers and designers who may not have technical expertise in XAI [2]. UX researchers and designers may struggle to select appropriate explanation methods, understand the limitations and biases of different techniques, or anticipate how users with varying levels of expertise will interpret the explanations, all of which are essential for designing trustworthy and user-centered XAI experiences. Without a structured approach, UX professionals struggle to translate AI explanations into user-friendly experiences, resulting in explanations that are ineffective, confusing, or ignored by users [3].

The need for explainability in AI arises from the increasing integration of AI systems in high-stakes decision-making contexts such as healthcare, finance, and law enforcement [4]. This is because even with AI assistance, decision-makers are still ultimately responsible for the decisions they make [5]. Hence, decision-makers interacting with XAI systems require understandable and usable explanations to build trust and make informed choices. However, simply providing explanations does not guarantee improved understanding or better decision-making. Studies have shown that decision-makers often struggle to effectively engage with XAI outputs, hindering their ability to leverage AI insights for optimal outcomes [4,6,7]. The ability to effectively explain AI decisions has far-reaching implications, not only for individual decision-makers but also for the ethical and responsible deployment of AI systems in critical domains [20].

One of the main challenges UX researchers face when designing XAI interfaces is the technical complexity of AI explanations [2]. Many explanation methods, such as SHAP (Shapley Additive Explanations), LIME (Local Interpretable Model-agnostic Explanations), and counterfactual explanations, require a foundational understanding of machine learning principles [8]. These methods are mostly designed to data scientists and machine learning engineers to debug their models [9]. This technical barrier can make it difficult for UX professionals to effectively evaluate and present explanations in ways that align with user expectations and mental models [10]. As a result, explanations may be presented in ways that are either too complex or too simplistic, reducing their effectiveness.

Another challenge is the cognitive load placed on users when interacting with explanations. Users may not have the time or willingness to engage deeply with AI-generated explanations, leading to cognitive disengagement [21]. If explanations are perceived as too technical, irrelevant, or redundant, users may ignore them altogether [7]. This disengagement undermines the purpose of XAI and may even lead to negative user experiences, where users feel overwhelmed or frustrated by the additional information.

Moreover, the risk of misinterpretation and misuse of explanations further complicates XAI design. Even when explanations are available, users may misinterpret them due to cognitive biases, prior experiences, or a lack of contextual understanding [21]. For instance, users might develop misplaced trust in AI systems if explanations appear overly confident, or they might reject valid AI-driven insights if explanations do not align with their preconceived notions [22]. Misinterpretation can lead to erroneous decision-making, which is particularly problematic in critical applications such as healthcare diagnostics or financial risk assessment.

Beyond these challenges, HCI research can plays a crucial role in how users engage with explanations. Effective XAI interfaces should prioritize usability, ensuring that explanations are not only technically sound but also intuitive, accessible, and contextually relevant. Explanations should mimic human reasoning patterns where possible, making them feel natural and relatable [23]. Additionally, learnability is a key factor—users should be able to develop a better understanding of AI behavior over time, refining their mental models as they interact with explanations [7].

To address these challenges, we propose a structured framework that provides UX researchers with practical tools to design better XAI interfaces. Our framework consists of four plays, each containing specific cards that guide UX designers in navigating human-centered perspective behind XAI, and these include explanation types, cognitive disengagement factors, risks of misinterpretation, and principles of human-centered XAI interaction. In the following sections, we will explore related word, our method and the proposed XAI UXR playbook. This playbook will empower UX researchers with actionable insights to create more persuasive and user-friendly AI explanations, ultimately bridging the gap between technical AI research and real-world user experience needs.

## 2 Related work

Several research efforts have attempted to bridge the gap between technical XAI methods and user experience design. Prior work has explored frameworks, guidelines, and toolkits that aim to make AI explanations more accessible and useful for end users.

One prominent approach is the AI Explainability Guidelines proposed by Google's People + AI Research (PAIR) initiative [11]. These guidelines outline principles for designing AI explanations, emphasising aspects such as user-centeredness, trust calibration, and the importance of providing explanations that align with user goals. While these guidelines offer general principles, they lack a structured methodology that UX designers can follow systematically when developing XAI interfaces.

Another relevant effort is IBM's AI Explainability 360 (AIX360) toolkit [12], which provides a collection of interpretability methods to help developers and designers create explainable AI systems. AIX360 offers practical tools and resources, but its focus remains largely technical, requiring substantial machine learning expertise. UX researchers without a strong technical background may struggle to integrate these tools effectively into their design workflows.

A similar initiative, focusing on a human-centered design approach, is presented in [13]. This work outlines a three-component approach: domain analysis, requirements elicitation, and multi-modal interaction design, demonstrated through a case study designing explanations for a Clinical Decision Support System (CDSS) for child health. This case study provides a set of user requirements and design patterns for an explainable medical diagnosis system, emphasizing the importance of expert user involvement in XAI development and offering potential generic solutions for common XAI design challenges. While providing practical guidelines for UX researchers, this approach is specific to CDSS.

In addition to toolkits and guidelines, empirical studies have investigated how users perceive and interact with AI explanations. Research by Ehsan et al. [14] introduced the concept of "rationalization" in AI, where explanations are framed in ways that resemble human reasoning processes. Their work suggests that explanations should be more natural and conversational to enhance user trust and comprehension. Similarly, studies on cognitive biases in AI explanations [15] highlight how users' preconceptions and mental models influence their interpretation of AI outputs, reinforcing the need for a structured UX approach.

Despite these advancements, there remains a gap in providing UX researchers with practical, structured methodologies tailored to XAI. Existing approaches often require technical expertise or offer broad principles without actionable steps for implementation. Our proposed framework addresses this gap by introducing a set of structured UX plays and cards that help designers systematically approach XAI interface design. This framework complements existing research by offering a hands-on, user-centered methodology that enables UX researchers to create more effective and engaging explanations by using User Experience Research (UXR) Point of View [16]. In the next section, we introduce our framework in detail, discussing its components and how it can be applied to address the challenges outlined in the introduction.

## 3 Method

This section outlines the methodology used in this study, which is grounded in the User Experience Research Point of View (UXR PoV) playbook [16,17]. This approach focuses on the creation and application of play cards, which provide a comprehensive set of strategies and tools for UX researchers and practitioners. The UXR PoV playbook methodology emphasises interdisciplinary collaboration by integrating diverse perspectives from the design, technology, and business sectors. Through iterative research cycles, the playbook helps teams identify critical UX challenges, generate actionable insights, and implement solutions that align with human-centered design principles. This structured yet flexible framework enables UX researchers to continuously refine designs to ensure they meet user needs while aligning with stakeholder expectations.

To adapt the UXR PoV playbook to the context of XAI, we incorporated the PhD work of the first author [18]. This framework categorizes UX challenges in XAI into four main areas: *Choosing the explanation class, User Engagement and Retention, Cognitive Biases and Misuse, and Human-Centered XAI Principles*. The adaptation process began with a series of structured meetings involving a diverse team of experts from human-computer interaction (HCI), UX design, industry, and XAI. During these meetings, the team systematically reviewed the challenges identified in the PhD word and mapped them to relevant components of the UXR PoV playbook.

The adaptation process consisted of three main stages:

- **Challenge Identification:** The team identified specific human-XAI interaction challenges, utilizing the PhD work of the first author [18]. These challenges were categorized into the four identified themes.
- **Card Customization:** Existing play cards were adjusted, and new cards were created to address XAI-specific issues, such as cognitive overload, user distrust, and interpretability gaps, within the relevant categories.
- **Validation and Refinement:** The customized playbook was tested through mock scenarios. Insights gathered from these activities were used to refine the play cards, ensuring their relevance and effectiveness in real-world XAI design contexts.

As a result, four main plays were developed, each containing several cards addressing specific challenges or opportunities within UX in XAI (see Section 4 for results). Each play (P) targets a particular challenge or opportunity in UX research. Within each play, multiple cards represent specific issues in XAI UX. Each card has a dual-sided format: the front side highlights the issue, card type, and includes an insightful quote along with related cards, while the back side offers detailed guidance with 'Best Practices' for addressing the issue.

## 4　The XAI UXR Playbook – Plays and Cards

This section introduces the core plays and associated cards in the XAI UX) Playbook, designed to address the unique challenges faced in designing explainable AI systems and to bridge the gap between technical XAI and designing XAI interfaces. Each play focuses on specific aspects of user experience (UX) in XAI and provides actionable strategies to enhance the understanding, trust, and engagement of users with AI-generated explanations. The plays and their corresponding cards offer practical solutions to improve the effectiveness of AI explanations, ensuring they align with human-centered design principles and cognitive processes.

*P1: Choosing the Explanation Class*

One of the initial stages of designing XAI is determining the type of explanation that best suits the user's needs and expectations. This play introduces UX researchers to different explanation classes and highlights their potential impact on user understanding. The cards in this play help UX professionals differentiate between global and local explanations, as well as other key categories.

- **Explain the Model (Global):** Provides a high-level overview of how the AI system functions as a whole, could be useful for expert users and regulatory compliance.
- **Explain a Local Prediction (Local):** Focuses on explaining specific AI-generated decisions, making it more relevant to end-users who seek immediate clarity on particular outcomes.
- **Counterfactual Explanation:** This explanation answers "What if" questions to show how changes in input affect the AI's outcome (e.g., In a loan application system: "If your income had been $10,000 higher, your loan would have been approved." This explains how a change in income would have altered the decision)
- **Example-based Explanation:** Uses past cases or similar examples to justify AI predictions, which can be particularly useful for users who relate better to real-world analogies.

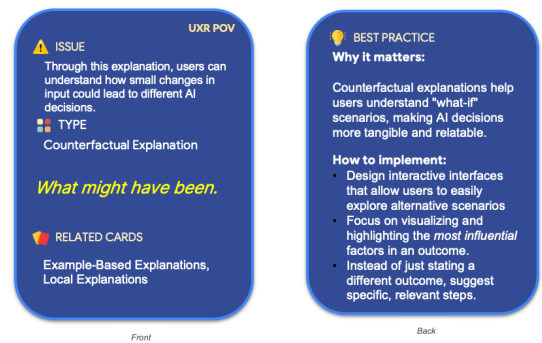

*P2: Understanding User Engagement and Retention*

Engaging users with AI explanations is a significant challenge, as users often disengage due to cognitive overload, lack of interest, or perceived irrelevance. This play identifies the key factors contributing to disengagement and provides strategies to enhance user retention and interaction with AI explanations.

- **Lack of Curiosity:** Users may not explore explanations if they do not perceive any immediate benefit. Solutions include making explanations visually engaging or interactive.
- **Perceived Goal Impediment:** If explanations disrupt users' primary tasks, they may be ignored. Designers should integrate explanations seamlessly into workflows.
- **Redundant Information:** Repetitive or overly simplistic explanations can frustrate users. Customizable or adaptive explanations can address this issue.
- **Perceived Complexity:** Users may avoid explanations if they appear too technical or overwhelming. Using layman's terms and progressive disclosure can improve accessibility.
- **Lack of Context:** Explanations that lack situational relevance may be dismissed. Providing contextualized justifications tailored to user intent can enhance engagement.

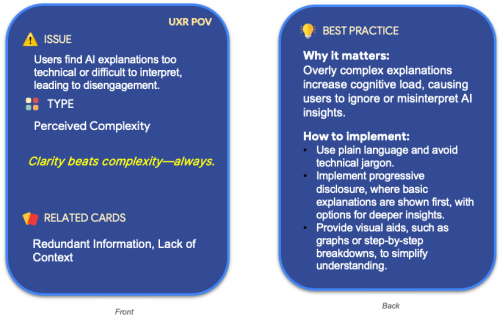

*P3: Acknowledging Cognitive Biases and Misuse*

Even when users engage with AI explanations, they may misinterpret or misuse them due to inherent cognitive biases and behaviors. This play addresses these biases, providing strategies to mitigate the risks of incorrect conclusions or misplaced trust in AI-generated explanations.

- **Misinterpretation:** Users may misunderstand technical details or draw incorrect conclusions. Clarity, consistency, and visual aids can minimize misinterpretation.
- **Mistrust:** Users may be skeptical of AI explanations, especially if they conflict with personal beliefs. Transparency and verifiability can help build confidence.
- **Confirmatory Search:** Users may selectively seek explanations that align with their preconceived notions. Encouraging diverse perspectives and presenting counterfactuals can counteract this bias.
- **Rush Understanding:** Users may skim explanations without fully processing them, leading to flawed decisions. Strategies such as progressive disclosure and interactive explanations can encourage deeper engagement.
- **Habit Formation:** Repeated exposure to explanations may lead to automatic, thoughtless acceptance. Rotating explanation styles and emphasizing critical reflection can prevent complacency.

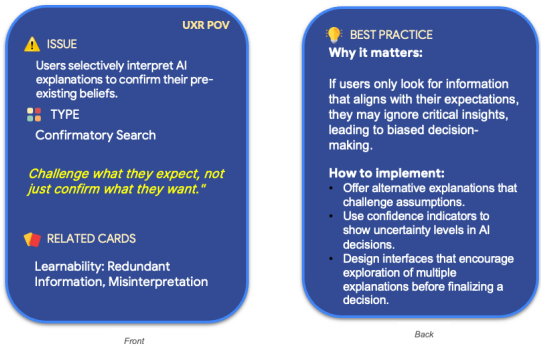

*P4: Creating User-Friendly and Intuitive XAI*

This play focuses on designing AI explanations that are easy to understand, relatable, and aligned with human cognitive processes, ensuring they are intuitive and accessible for all users. The goal is to foster long-term understanding and trust in AI by making the interaction as natural and user-centric as possible.

- **Usable Explanations:** Explanations should be easy to understand, well-structured, and presented in a digestible format.
- **Human-like Explanations:** AI-generated explanations should mimic human reasoning styles where appropriate, making them feel more relatable and intuitive.
- **Learnability of Explanations:** Users should be able to develop a better understanding of AI decision-making over time. This can be achieved through progressive learning approaches, where explanations evolve based on user interaction and familiarity.

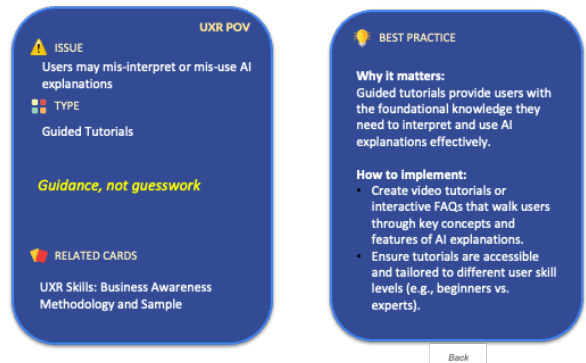

## 5 Initial Evaluation

To evaluate the effectiveness of the proposed XAI UXR Playbook, we conducted a study in which we used Generative AI (Gen AI) as an initial evaluator before engaging human UX researchers. This approach enabled us to rapidly identify usability challenges, inconsistencies, and areas for refinement. This evaluation approach has been used in similar studies and showed that using Gen AI can complement user studies [19]. The aim was to refine the playbook before subjecting it to more extensive human-centered evaluation with UX designers. The evaluation aimed to assess how well the playbook's cards guide UX research in designing AI explanations.

We tested the playbook's cards across multiple Gen AI models, including OpenAI's GPT-4, Google's Gemini, and Meta's Llama. Each model was prompted with various cards, instructing them to apply the guidance provided to AI explanation design. The results showed that the structured nature of the cards was effective in guiding the AI models, particularly in generating user-friendly and interpretable explanations. In particular, cards from *P2: Understanding User Engagement and Retention*—such as *"Perceived Complexity"* and *"Lack of Context"*—helped AI models generate explanations that were more concise and contextually relevant. For instance, when asked to create an explanation for a loan approval decision, models following the *"Lack of Context"* card produced explanations that incorporated user-specific financial history rather than generic statements. These findings indicate that the playbook successfully encapsulates core UX principles that can be readily understood and implemented by AI models.

However, we also observed notable inconsistencies in Gen AI outputs. In particular, concerning *P3: Acknowledging Cognitive Biases and Misuse*. For instance, when applying the "*Rush Understanding"* card—designed to mitigate users' tendency to skim explanations without fully processing them—AI models demonstrated varying levels of effectiveness. When tasked with explaining why a loan application was denied, one AI model effectively applied progressive disclosure by first providing a concise summary (e.g., "Your loan was denied due to insufficient income") and then offering expandable sections for users who wanted more details. This approach encouraged deeper engagement without overwhelming the user. However, another AI model generated a single, dense paragraph with all relevant details at once, assuming the user would read it thoroughly. In usability testing, such explanations often lead to cognitive overload, increasing the risk of misinterpretation or complete disengagement. This finding highlights the importance of structuring explanations in a way that accommodates users' natural reading behaviors, ensuring they process critical information rather than skimming past it.

Overall, this initial evaluation demonstrated that the playbook provides structured and effective UX guidance for AI explanation design. However, the inconsistencies observed in AI-generated explanations suggest a need for clearer card instructions to ensure consistent interpretation across different AI and human users. Additionally, enhancing the example-based guidance on each card can provide clearer illustrations of best practices, helping users effectively translate UX principles into actionable design strategies. Another critical refinement is the adaptation of cards to accommodate varying levels of user expertise, ensuring that both novice and experienced UX professionals can apply them effectively. Moreover, integrating more structured recommendations for interactive and adaptive explanation techniques can further support user engagement, particularly in addressing cognitive biases and optimizing information retention.

## 6 Conclusion

The evaluation of the XAI UXR Playbook highlights its strengths in providing structured guidance for UX researchers and its potential for improving explainability in AI systems. Leveraging Gen AI as an initial evaluation tool proved to be an efficient method for identifying usability issues, refining content, and streamlining the research process before involving human UX professionals. While the cards facilitated AI-generated UX insights, inconsistencies in interpretation and limitations in contextual understanding reinforce the importance of human oversight. Our future research aims to focus on refining card instructions, tailoring them for domain-specific applications, and integrating interactive elements to enhance usability.

## REFERENCES

[1] Arrieta, A.B., Díaz-Rodríguez, N., Del Ser, J., Bennetot, A., Tabik, S., Barbado, A., García, S., Gil-López, S., Molina, D., Benjamins, R. and Chatila, R., 2020. Explainable Artificial Intelligence (XAI): Concepts, taxonomies, opportunities and challenges toward responsible AI. Information fusion, 58, pp.82-115.
[2] Naiseh, M., Simkute, A., Zieni, B., Jiang, N. and Ali, R., 2024. C-XAI: A conceptual framework for designing XAI tools that support trust calibration. Journal of Responsible Technology, 17, p.100076.
[3] Ferreira, J.J. and Monteiro, M.S., 2020. What are people doing about XAI user experience? A survey on AI explainability research and practice. In Design, User Experience, and Usability. Design for Contemporary Interactive Environments: 9th International Conference, DUXU 2020, Held as Part of the 22nd HCI International Conference, HCII 2020, Copenhagen, Denmark, July 19–24, 2020, Proceedings, Part II 22 (pp. 56-73). Springer International Publishing.
[4] Naiseh, M., Al-Thani, D., Jiang, N. and Ali, R., 2023. How the different explanation classes impact trust calibration: The case of clinical decision support systems. International Journal of Human-Computer Studies, 169, p.102941.
[5] Naiseh, M., Bentley, C. and Ramchurn, S.D., 2022, March. Trustworthy autonomous systems (TAS): Engaging TAS experts in curriculum design. In 2022 IEEE global engineering education conference (EDUCON) (pp. 901-905). IEEE.
[6] Cabitza, F., Fregosi, C., Campagner, A. and Natali, C., 2024, July. Explanations considered harmful: the impact of misleading explanations on accuracy in hybrid human-ai decision making. In World conference on explainable artificial intelligence (pp. 255-269). Cham: Springer Nature Switzerland.
[7] Simkute, A., Surana, A., Luger, E., Evans, M. and Jones, R., 2022, October. XAI for learning: Narrowing down the digital divide between "new" and "old" experts. In Adjunct Proceedings of the 2022 Nordic Human-Computer Interaction Conference (pp. 1-6).
[8] Sokol, K. and Flach, P., 2020, January. Explainability fact sheets: a framework for systematic assessment of explainable approaches. In Proceedings of the 2020 conference on fairness, accountability, and transparency (pp. 56-67).
[9] Vilone, G. and Longo, L., 2021. Notions of explainability and evaluation approaches for explainable artificial intelligence. Information Fusion, 76, pp.89-106.
[10] Buçinca, Z., Malaya, M.B. and Gajos, K.Z., 2021. To trust or to think: cognitive forcing functions can reduce overreliance on AI in AI-assisted decision-making. Proceedings of the ACM on Human-computer Interaction, 5(CSCW1), pp.1-21.
[11] Explainability + Trust. (n.d.). People + AI Guidebook. https://pair.withgoogle.com/chapter/explainability-trust/
[12] Mojsilovic, A. (2023, November 17). Introducing AI Explainability 360. IBM Research. https://research.ibm.com/blog/ai-explainability-360
[13] Schoonderwoerd, T.A., Jorritsma, W., Neerincx, M.A. and Van Den Bosch, K., 2021. Human-centered XAI: Developing design patterns for explanations of clinical decision support systems. International Journal of Human-Computer Studies, 154, p.102684.
[14] Ehsan, U., Harrison, B., Chan, L. and Riedl, M.O., 2018, December. Rationalization: A neural machine translation approach to generating natural language explanations. In Proceedings of the 2018 AAAI/ACM Conference on AI, Ethics, and Society (pp. 81-87).
[15] Kaur, H., Conrad, M.R., Rule, D., Lampe, C. and Gilbert, E., 2024. Interpretability Gone Bad: The Role of Bounded Rationality in How Practitioners Understand Machine Learning. Proceedings of the ACM on Human-Computer Interaction, 8(CSCW1), pp.1-34.
[16] Dogan, H., Giff, S. and Barsoum, R., 2024, May. User Experience Research: Point of View Playbook. In Extended Abstracts of the CHI Conference on Human Factors in Computing Systems (pp. 1-7).
[17] UXR POV PlayBook. (n.d.). UXR POV Playbook. https://www.uxrpovplaybook.com/
[18] Naiseh, M., 2021. C-XAI: Design Method for Explainable AI Interfaces to Enhance Trust Calibration (Doctoral dissertation, Bournemouth University).
[19] Bombassei De Bona, F., Dominici, G., Miller, T., Langheinrich, M. and Gjoreski, M., 2024. Evaluating Explanations Through LLMs: Beyond Traditional User Studies. arXiv e-prints, pp.arXiv-2410.
[20] Naiseh, M., 2024. Social eXplainable AI (Social XAI): Towards expanding the social benefits of XAI. In The impact of artificial intelligence on societies: Understanding attitude formation towards AI (pp. 169-178). Cham: Springer Nature Switzerland.
[21] Naiseh, M., Cemiloglu, D., Al Thani, D., Jiang, N. and Ali, R., 2021. Explainable recommendations and calibrated trust: two systematic user errors. Computer, 54(10), pp.28-37.
[22] Cabitza, F., Natali, C., Famiglini, L., Campagner, A., Caccavella, V. and Gallazzi, E., 2024. Never tell me the odds: Investigating pro-hoc explanations in medical decision making. Artificial intelligence in medicine, 150, p.102819.
[23] Miller, T., 2019. Explanation in artificial intelligence: Insights from the social sciences. Artificial intelligence, 267, pp.1-38.